\documentstyle[11pt,aasms4,flushrt,epsf,rotate]{article}

\def\bt{\begin{tabbing}}
\def\et{\end{tabbing}}
\def\beq#1{\begin{equation}\label{#1}}
\def\eeq{\end{equation}}

\def\khat{\widehat{\bf k}}
\def\rhat{\widehat{\bf r}}

\def\etal{{\frenchspacing\it et al.}}
\def\ie{{\frenchspacing\it i.e.}}
\def\eg{{\frenchspacing\it e.g.}}

\def\rms{{\frenchspacing r.m.s.}}

\def\kt{k_\perp}
\def\FWHM{{\rm FWHM}}

\def\spose#1{\hbox to 0pt{#1\hss}}
\def\simlt{\mathrel{\spose{\lower 3pt\hbox{$\mathchar"218$}}
     \raise 2.0pt\hbox{$\mathchar"13C$}}}
\def\simgt{\mathrel{\spose{\lower 3pt\hbox{$\mathchar"218$}}
     \raise 2.0pt\hbox{$\mathchar"13E$}}}      

\slugcomment{Published in {\it{ApJ}}, 468:457 (1996).}

\begin{document}
\title{CAN THE LACK OF SYMMETRY IN THE $COBE$/DMR MAPS CONSTRAIN
       THE TOPOLOGY OF THE UNIVERSE?}

\bigskip

\author{Ang\'{e}lica de Oliveira-Costa$^{1,2}$, George F. Smoot$^{1}$ \&
        Alexei A. Starobinsky$^{3}$}

\bigskip

\affil{$^1$Lawrence Berkeley Laboratory, Space Sciences 
Laboratory \& Center for Particle Astrophysics, Building 50-205, University 
of California, Berkeley, CA 94720; angelica@cosmos.lbl.gov}
 
\affil{$^2$Instituto Nacional de Pesquisas Espaciais (INPE), 
Astrophysics Division, S\~{a}o Jos\'{e} dos Campos, S\~{a}o Paulo 12227-010, 
Brazil.}

\affil{$^3$Russian Academy of Sciences \& Landau Institute for 
Theoretical Physics, Kosygina St. 2, Moscow 117334, Russia.}

\begin{abstract}
Although the cubic $T^3$ ``small universe" has been ruled out by $COBE$/DMR 
results as an interesting cosmological model, we still have the possibility of 
living in a universe with a more anisotropic topology such as a rectangular 
$T^3$ ``small universe" with one or two of its dimensions significantly smaller 
than the present horizon (which we refer to as $T^1$- and $T^2$-models, 
respectively).
In order to rule out these anisotropic topologies as well, we apply a new data 
analysis method that searches for the specific kind of symmetries that these 
models should produce. 
We find that the 4 year $COBE$/DMR data set a lower limit on the smallest cell 
size for $T^1$- and $T^2$-models of 3000$h^{-1}$Mpc, at 95\% confidence, 
for a scale invariant power spectrum ($n$=1). 
These results imply that {\it all} toroidal universes (cubes and rectangles) 
are ruled out as interesting cosmological models. 
\end{abstract}

\keywords{cosmic microwave background, large-scale structure of universe.}

\section{INTRODUCTION}

In the past few years, mainly after the discovery of CMB anisotropies by 
$COBE$/DMR (Smoot {\etal} 1992), 
the study of the topology of the universe has become an important problem for 
cosmologists and some hypotheses, such as the ``small universe" model 
(see {\eg} Ellis and Schreiber 1986), have received considerable attention. 
From the theoretical point of view, it is possible to have quantum creation of 
the universe with a multiply-connected topology (Zel'dovich and Starobinsky 
1984). From the observational side, this model has been used to explain the 
``observed" periodicity in the distributions of quasars (Fang and Sato 1985) 
and galaxies (Broadhurst {\etal} 1990).

Almost all work on ``small universes" has been limited to the case 
where the spatial sections form a rectangular basic cell with sides 
$L_{x}, L_{y}, L_{z}$ and with opposite faces topologically connected, a 
topology known as toroidal. 
The three-dimensional cubic torus $T^3$ is the simplest model among all 
possible multiply-connected topologies, in which all three sides have the same 
size $L \equiv L_{x} = L_{y} = L_{z}$. 
In spite of the fact that cubic $T^3$-model has been ruled out by $COBE$ 
results (Sokolov 1993; Starobinsky 1993, hereafter S93; Stevens {\etal} 1993; 
Jing and Fang 1994; de Oliveira-Costa and Smoot 1995, hereafter dOCS95), the 
possibility that we live in a universe with a more anisotropic topology, such 
as a rectangular torus $T^3$, is an open problem that has not been ruled out 
yet.
For instance, if the toroidal model is not a cube, but a rectangle with sides 
$L_{x} \neq L_{y} \neq L_{z}$ and with one or two of its dimensions 
significantly smaller than the horizon $R_H$ ($\equiv 2cH_{0}^{-1}$), this 
small rectangular universe cannot be completely excluded by any of the previous 
analyses: constraints from the DMR data merely require that at least one of 
the sides of the cell be larger than $R_H$. 

As pointed out by S93 and Fang (1993), if the rectangular $T^3$-universe has 
one of the cell sizes smaller than the horizon and the other two cell sizes are
of the order of or larger than the horizon (for instance, $L_x \ll R_H$ and 
$L_y,L_z \simgt R_H$), the large scale CMB pattern shows the existence of a 
symmetry plane, and 
if it has two cell sizes smaller than the horizon and the third cell size is 
of the order of or larger than the horizon (for instance, $L_x,L_y \ll R_H$ and 
$L_z \simgt R_H$), the CMB pattern shows the existence of a symmetry axis. 
We call the former case a $T^1$-model because the spatial topology of the 
universe becomes just $T^1$ in the limit $L_y,L_z \to \infty$ with $L_x$ being
fixed. The later case is denoted a $T^2$-model for the same reason (the 
corresponding limit is $L_z \to \infty$ with $L_x,L_y$ being fixed). 
It is clear (and our calculations confirm it) that dependence of CMB
fluctuations on any cell size is very small once it exceeds the horizon.
In previous work (dOCS95), we computed the full covariance matrix for all 
multipole components and used a $\chi^2$-technique to place a lower limit on
the cell size $L$ of the cubic $T^3$-models. However, we cannot apply this same 
approach to study the $T^1$- and $T^2$-universes. As we explain in the next 
section, the observed power spectrum of these models depends not only on the 
cell size but also strongly on the cell orientation relative to the Galaxy cut. 

Our goal is to show that the $COBE$/DMR maps have the ability to discriminate 
and rule out $T^1$- and $T^2$-models. 
We use a different approach to study these models in which we constrain their 
sizes by looking for the symmetries that they would produce in the CMB,
obtaining strong constraints from the 4 year $COBE$/DMR data.

\section{SYMMETRIES IN THE CMB DUE TO TOPOLOGY}

If the density fluctuations are adiabatic and the Universe is spatially
flat, the Sachs-Wolfe fluctuations in the CMB are given by
\beq{Sachs_Wolfe} \frac{\delta T}{T} (\theta,\phi) = - \frac{1}{2} 
\frac{H_{0}^{2}}{c^{2}} \sum_{\bf k} \frac{\delta_{\bf k}}{{k}^{2}} 
e^{i{\bf k} \cdot {\bf r}} \eeq
(Peebles, 1982), where {\bf r} is the vector with length 
$R_H \equiv 2c H_{0}^{-1}$ that points in the direction of observation 
($\theta,\phi$), $H_{0}$ is the Hubble constant (written here as 
$100h$~km s$^{-1}$ Mpc$^{-1}$) and $\delta_{\bf k}$ is the density fluctuation 
in Fourier space, with the sum taken over all wave numbers {\bf k}. 
Here we neglect the difference between the horizon surface and the surface of
the last scattering, which is justified for $l <$ 30. 

It is customary to expand the CMB fluctuations in spherical harmonics
\beq{deltaT} \frac{\delta T}{T}(\theta,\phi) = \sum_{l=0}^{\infty}
\sum_{m=-l}^{l} a_{lm} Y_{lm} (\rhat), \eeq
where $a_{lm}$ are the spherical harmonic coefficients and $\rhat$ is 
the unit vector in direction ${\bf r}$. The coefficients $a_{lm}$ are given by
\beq{alm} a_{lm} = -2\pi i^{l} \frac{H_{0}^{2}}{c^{2}} \sum_{\bf k}  
\frac{\delta_{\bf k}}{k^{2}} j_{l}(kR_H) Y_{lm}^{*}(\khat), \eeq
where $j_{l}$ are spherical Bessel functions of order $l$. If we assume that
the CMB fluctuations $\delta T/T$ are a Gaussian random field, the coefficients
$a_{lm}$ are Gaussian random variables with zero mean and covariance matrix 
\beq{var_alm} \langle a_{lm}^{*}a_{l'm'} \rangle 
\propto \sum_{\bf k}  
\frac{ | \delta_{\bf k}|^2 }{k^{4}} j_{l}(kR_H) j_{l'}(kR_H) Y_{lm}(\khat)
Y_{l'm'}^{*}(\khat). \eeq
In a Euclidean topology the universe is isotropic, the sum in (\ref{var_alm}) 
is replaced by an integral and the power spectrum $C_{l}$ is related to the
coefficients $a_{lm}$ by
\beq{Cl_alm} \langle a_{lm}^{*}a_{l'm'} \rangle 
\equiv C_{l} \delta_{ll'}\delta_{mm'} \eeq
(see {\eg} Bond and Efsthatiou 1987). 
However, in a toroidal universe this is not the case. In this model, only wave 
numbers that are harmonics of the cell size are allowed. We have a discrete 
{\bf k} spectrum
\beq{kcut} {\bf k} = \frac{2 \pi}{R_H} \left(  \frac{p_x}{R_x}, \frac{p_y}{R_y},
\frac{p_z}{R_z} \right) \eeq
(Zel'dovich 1973; Fang and Houjun 1987), where $p_x$, $p_y$ and $p_z$ are 
integers and $R_x \equiv L_x/R_H$, $R_y \equiv L_y/R_H$ and $R_z \equiv 
L_z/R_H$.  

In previous work (dOCS95), we set limits on the cubic $T^3$-models assuming 
that, for a given cell size, the quantity $\hat{C}_l \equiv \frac{1}{2l+1} \sum 
|a_{lm}|^2$ was fairly independent of the cell orientation, even with a
20$^{\circ}$ Galaxy cut. In other 
words, if $\hat{C}_l$ is almost independent of the cell orientation, we can 
make the approximation that all cell orientations for that given cell size can 
be simultaneously ruled out by a $\chi^2$-test on the $\hat{C}_l$ 
coefficients and, in that way, test our model just considering changes in the 
cell size $L$. 
However, in the case of  more strongly anisotropic cell configurations such as 
$T^1$- and $T^2$-models, the quantity $\hat{C}_l$ does depend on the cell 
orientation and the $\chi^2$-test on the power spectrum cannot be used 
anymore. If we try to apply the power spectrum method to these 
models, it will require testing a six parameter family of models with three 
parameters corresponding to the cell orientation in addition to the cell sizes 
$L_x$, $L_y$ and $L_z$.
          
In order to illustrate these anisotropic cell configurations, we plot a
realization of two extreme cases: a $T^1$-universe (Figure~1A, upper 
left) with dimensions ($R_x,R_y,R_z$) = (3,3,0.3) and a $T^2$-universe 
(Figure~1B, upper right) with dimensions ($R_x,R_y,R_z$) = (0.3,0.3,3). 
Both models are plotted in galactic coordinates and have a scale invariant
power spectrum ($n$=1). From equations 
(\ref{Sachs_Wolfe}) and (\ref{kcut}), we see that when one of the cell sizes 
is smaller than the horizon ($T^1$-models), the temperatures $\delta T/T$ are 
almost independent of this coordinate. For instance if $R_z \ll $1, the values 
of $\delta T/T$ are almost independent of the $z$-coordinate, \ie, the values of $\delta T/T$ are symmetric about the plane formed by the $x$ and $y$-axes. 
This happens because of the factor $\delta_{\bf k}/k^2$ in equation 
(\ref{Sachs_Wolfe}). If we assume a power-law power spectrum with $n$=1, the
{\rms} value of this factor scales as $k^{-3/2}$, so that most of the 
contribution to the sum comes from small $k$-values. If $R_z \ll $1, the 
term $p_z/R_z$ in equation (\ref{kcut}) will be much greater than unity when 
$p_z \ne$ 0, so the term with $p_z =$ 0 will dominate the sum. 
Since this term is independent of the $z$-coordinate, the entire sum will be 
approximatelly independent of $z$. In the same way, if two cell sizes are 
smaller than the horizon ($T^2$-models), the temperatures $\delta T/T$ are 
aproximately independent of these coordinates. For instance if $R_x, R_y \ll 
$1, the values of $\delta T/T$ are almost independent of both $x$ and $y$, 
\ie, the values of $\delta T/T$ are almost constant along rings around the 
$z$-axis. 

The results above remain valid for a much broader range of $n$-values (actually,
$n <$ 3). Thus, the following analysis is applicable in {\it any} other large
scale CMB experiment as well as one-degree CMB experiments. 
Although the existence of these symmetry patterns in the large scale
fluctuations $\delta T/T$ do not depend on the assumptions of gaussian
statistics and absence of correlation between multipoles (see S93), we use both of these standard assumptions in this analysis.

The analysis of $T^1$- and $T^2$-models is not an easy task, since there are
infinitely many combinations of different cell sizes and cell orientations. In 
order to keep our analysis simple, we wish to adopt a statistic that 
is independent of cell orientation. In addition, we want a statistic that 
is precisely sensitive to the type of symmetries that small universes 
produce, so that it can rule out as many false models as possible. Finally, we 
would like to have a statistic that is easy to compute and that produces
results that are easy to interpret.
Having these criteria in mind, we choose a statistic in which 
we calculate the function $S(\hat{\bf n}_{i})$ defined by
\beq{Smin_plane} S(\hat{\bf n}_{i}) \equiv \frac{1}{N_{pix}} 
\sum_{j=1}^{N_{pix}} \frac{ [ \frac{\delta T}{T}(\hat{\bf n}_j) - 
\frac{\delta T}{T}(\hat{\bf n}_{ij}) ]^{2}}{ \sigma(\hat{\bf n}_j)^{2} + 
\sigma(\hat{\bf n}_{ij})^{2}}, \eeq 
where $N_{pix}$ is the number of pixels that remain in the map after the Galaxy
cut have taken place, $\hat{\bf n}_{ij}$ denotes the reflection of 
$\hat{\bf n}_{j}$ in the plane whose normal is $\hat{\bf n}_{i}$, \ie, 
\beq{normal} \hat{\bf n}_{ij} = \hat{\bf n}_{j} - 2 (\hat{\bf n}_{i} \cdot
\hat{\bf n}_{j}) \hat{\bf n}_{i} \eeq
and $\sigma(\hat{\bf n}_{j})$ and $\sigma(\hat{\bf n}_{ij})$ are the {\rms} 
errors associated with the pixels in the directions $\hat{\bf n}_{j}$ and 
$\hat{\bf n}_{ij}$. 
$S(\hat{\bf n}_{i})$ is a measure of how much reflection symmetry there is in 
the mirror plane perpendicular to $\hat{\bf n}_{i}$. The more perfect the 
symmetry is, the smaller $S(\hat{\bf n}_{i})$ will be. 
When we calculate $S(\hat{\bf n}_{i})$ for all 6144 pixels at the positions 
$\hat{\bf n}_{i}$, we obtain a sky map that we refer to as an $S$-map. This 
sky map is a useful visualization tool and gives intuitive understanding of
how the statistic  $S(\hat{\bf n}_{i})$ works. 

In order to better understand $S(\hat{\bf n}_{i})$, we first consider the
simple model of a $T^1$-universe with $R_z \ll $1.
For this specific model, the values of $\delta T/T$ are almost independent 
of the $z$-coordinate, so we have almost perfect mirror symmetry about the 
$xy$-plane or, in spherical coordinates, $\delta T/T (\theta,\phi) \approx 
\delta T/T (\pi-\theta,\phi)$. 
When $\hat{\bf n}_{i}$ points in the direction of the smallest cell size (\ie,
$z$-direction), we have $S(\hat{\bf n}_{i}) \approx$ 1; otherwise, 
$S(\hat{\bf n}_{i}) >$ 1. 
An $S$-map for a $T^1$-model ($R_x,R_y,R_z$) = (3,3,0.3) can be seen in 
Figure~1C (lower left). Notice in this plot that the direction in which 
the cell is smallest can be easily identified by two ``dark spots" at 
$\hat{\bf n}_{i} \approx \hat{\bf z}$ and $\hat{\bf n}_{i} \approx - \hat{\bf 
z}$. For $T^2$-models, the only difference will be that in the place of the two 
``dark spots", we have a ``dark ring" structure in the plane formed
by the two small directions. See Figure~1D (lower right), an $S$-map of the 
$T^2$-model ($R_x,R_y,R_z$) = (0.3,0.3,3).

From these two $S$-maps, we can infer two important properties: first, 
the direction in which the $S$-map takes its minimum value, denoted 
$S_{\circ}$, is the direction in which the universe is small. For a large
universe such as (3,3,3), the $S_{\circ}$-directions obtained from different 
realizations are randomly distributed in the sky. 
Secondly, the distribution of $S_{\circ}$-values changes with the cell size, 
\ie, as the universe becomes smaller, the values of $S_{\circ}$ decrease.
From the definition of the $S$-map, it is easy to see that the value of 
$S_{\circ}$ is independent of the cell orientation. In other words, if we rotate
the cell, we will just be rotating the $S$-map, leaving its minimum value 
$S_{\circ}$ unchanged.

A value of $S_{\circ}$ from a particular realization 
of a stochastic cosmological perturbation differs from
the expectation value $\langle S_{\circ}\rangle$ due to cosmic variance. 
This comes from the non-symmetric part of $\delta T/T$ fluctuations 
produced by perturbation modes with $p_x + p_y + p_z \neq 0$. 
For these modes, the main contribution to $S(\hat{\bf n}_i)$ in 
(\ref{Smin_plane}) is made by the terms in the sum for which 
$\hat{\bf n}_j$ and $\hat{\bf n}_{ij}$ are widely separated, so that we can
neglect their cross-correlation. Since 
$S \approx {1\over\sigma^2}
\left\langle \left( \frac{\delta T}{T} \right)^2 \right\rangle_{ns}$, 
where 
$\left\langle \left( \frac{\delta T}{T} \right)^2 \right\rangle^{1/2}_{ns}$ is 
the r.m.s. value of the
non-symmetric part of $\delta T/T$ and $\sigma$ is the
r.m.s. noise, we have that the cosmic variance is 
$\Delta S \equiv \sqrt{\langle S^2\rangle - \langle S\rangle^2}
\approx {1\over\sigma^2}
\sqrt{\frac{2}{2l+1}} 
\left\langle \left( \frac{\delta T}{T} \right)^2 \right\rangle_{ns}
\approx 0.2S$. 
Here $l \approx 15$ is the characteristic multipole for $COBE$ data
(the inverse angular correlation radius) and
$\left\langle\left(\frac{\delta T}{T}\right)^2\right\rangle_{ns}^{1/2}
\leq 
\sigma_{7^{\circ}}$. 
We shall confirm this rough estimate in more details below; see the
behavior of curves for the cumulative probability distribution of 
$S_{\circ}$ in Figure~2.

In summary, our statistic $S_{\circ}$ has all the properties that we desire: it 
quantifies the ``smallness" of a sky map in a single number, it is independent
of the cell orientation, and it is easy to work with and to interpret. 

From here on, we will present our results in terms of the cell sizes $R_x$, 
$R_y$ and $R_z$, usually sorted as $R_x \le R_y \le R_z$. We 
remind the reader that the results are identical for all six permutations of 
$R_x$, $R_y$ and $R_z$.

\section{DATA ANALYSIS}

We rewrite the exponential in equation (\ref{Sachs_Wolfe}) as
\beq{exponential} e^{i{\bf k} \cdot {\bf r}} = \cos{{\bf k} \cdot {\bf r}} 
+ i \sin{{\bf k} \cdot {\bf r}} = \cos(2 \pi \gamma) + i \sin(2 \pi \gamma),
\eeq
where ${\bf k}$ is given by (\ref{kcut}), ${\bf r}$ is the vector with lenght
$R_H \equiv 2cH_{0}^{-1}$ and 
$\gamma = \left( \frac{p_x}{R_x} x + \frac{p_y}{R_y} y + \frac{p_z}{R_z} 
z \right)$. 
If the density fluctuation in Fourier space $\delta_{\bf k}$ has random phases,
we have $\delta_{\bf k} = N(g_1 + ig_2)$, so that 
$\langle |\delta_{\bf k}|^2 \rangle = \langle |g_1 + ig_2|^2 N^2 \rangle = 
2N^2$, where $N$ is a constant and $g_{1}$ and $g_{2}$ are two 
independent Gaussian random variables with zero mean and unit variance. 
Assuming a power law power spectrum with shape 
$P(k) \equiv \langle |\delta_{\bf k}|^2 \rangle = Ak^n$, where $A$ is the 
amplitude of scalar perturbations and $n$ the spectral index, we have 
$N = \sqrt{\frac{A}{2}} k^{n/2}$, so that the {\rms} of the term $\delta_{\bf k}/k^2$ in (\ref{Sachs_Wolfe}) is given by
\beq{deltak_k} \frac{ \langle |\delta_{\bf k}|^2 \rangle^{1/2} }{k^2} 
\propto \alpha^{\frac{n-4}{4}}, \eeq
where $\alpha \equiv \left( \frac{p_x}{R_x} \right)^2 + \left( \frac{p_y}{R_y} 
\right)^2 + \left( \frac{p_z}{R_z} \right)^2 \propto k^2$.
Substituting (\ref{exponential}) and (\ref{deltak_k}) into (\ref{Sachs_Wolfe}), 
we can construct simulated skies by calculating
\beq{new_SW} \frac{\delta T}{T} (\theta,\phi) \propto
\sum_{p_x, p_y, p_z} \left[ g_1 \cos(2 \pi \gamma) + g_2 \sin(2 \pi \gamma) 
\right] \alpha^{\frac{n-4}{4}}. \eeq
                                                 
Since the cubic $T^3$ case has already been ruled out as an interesting 
cosmological model (see {\eg} dOCS95), we restrict our analysis here to the 
$T^1$ and $T^2$ cases for $n$=1. This is a two parameter family of models 
specified by $R_x$ and $R_y$, with $R_z=\infty$. For numerical convenience, we 
set $R_z$=3 instead, as this is found to give virtually the same results as 
$R_z=\infty$. We adopt $n$=1, as we found that ``small universe" models with 
different $n$-values are even more inconsistent with the observed 
data.

The large scale fluctuations observed on the celestial sphere by a CMB 
experiment can be modeled as being the fluctuations given in (\ref{new_SW}) 
multiplied by an experimental beam function 
\beq{smoothing} e^{-(R_H \theta \kt)^{2}/2}, \eeq
where $\kt$ is the length of the ${\bf k}$-component perpendicular to the 
line of sight and $\theta$ is the width of the Gaussian beam given by
$\theta = \FWHM / \sqrt{8 \ln{2}} \approx 0.43~\FWHM$, where FWHM is the 
full width of the beam at its half maximum. 
We make the approximation that the sky area covered by the beam is flat 
(this is equivalent to smoothing in the plane perpendicular to the line of 
sight $\hat{\bf r}$). 
Since $\kt \equiv | \hat{\bf r}  \times \hat{\bf k} |$, we have that $\kt^2 
= k^2 - (\hat{\bf r} \cdot \hat{\bf k})^2$. 
We use FWHM = 7$^{\circ}$ in our simulations, which is the FWHM of the 
$COBE$/DMR beam.

In the real sky map, we do not have complete sky coverage. Because of the 
uncertainty in Galaxy emission, we are forced to remove all pixels less than 
$20^{\circ}$ below and above the Galaxy plane, which represents a loss of almost 
34\% of all pixels. However, performing Monte Carlo simulations with and 
without the Galaxy cut, we find that the Galaxy cut does not change the final 
distribution of $S_{\circ}$ much. 
Due to the smaller data sample, the Galaxy cut weakens the lower limit on the 
cell size slightly (see {\eg} Scott {\etal} 1994). 

We model the noise $n_i$ at each pixel $i$ as independent Gaussian random 
variables with mean $\langle n_i \rangle =$0 and variance
$\langle n_in_j \rangle = \sigma_{ij} \delta_{ij}$ 
(Lineweaver {\etal} 1994), and add it to the temperature values 
given by (\ref{new_SW}). The level of noise in the DMR maps is a source of
serious concern in our analysis: high levels of noise can make it 
impossible to discriminate between the different topological models. In order 
to reduce the noise and increase the signal-to-noise ratio in the simulated skies 
and real data, we smooth both once more by 7$^{\circ}$ before calculating 
$S$-map, which corresponds to a total smearing of $\sqrt{ (7^{\circ})^2 + 
(7^{\circ})^2} \approx 10^{\circ}$. 

We generate our simulated skies as standard DMR maps with 6144 pixels for 
$n$=1, with a Galaxy cut of 20$^{\circ}$, FWHM = 7$^{\circ}$, the monopole and 
dipole removed, add noise and normalize to $\sigma_{7^{\circ}} = 
34.98\mu$K (the {\rms} value at 7$^{\circ}$ extracted from our DMR map, a 
4 year combined 53 plus 90 GHz map with monopole and dipole removed). 
Fixing a cell size, we construct a simulated sky according to (\ref{new_SW}), 
we smooth this once more by 7$^{\circ}$ and use the statistic defined
in (\ref{Smin_plane}) to obtain an $S$-map from which we extract its minimum 
value $S_{\circ}$. 
Repeating this procedure 1000 times, we obtain the probability distribution 
of $S_{\circ}$ for that fixed cell size. 
When we repeat this same procedure for different cell sizes, we are 
able to construct Figure~2. 

In Figure~2A (upper plot), we show the cumulative probability distribution of 
$S_{\circ}$ obtained from the Monte Carlo simulations for the cell sizes 
($R_x,R_y,R_z$) = (0.5,0.5,3), (0.6,0.6,3), (0.7,0.7,3) and (3,3,3). The 
horizontal lines indicate the confidence levels of 95\%, 90\% and 
68\% (from top to bottom). 
Comparing these curves with the value $S_{\circ}^{DMR}$ = 2.59 (represented in 
the plot by the vertical straight line), where $S_{\circ}^{DMR}$ is the 
$S_{\circ}$ value extracted from our data set, we conclude that $T^2$-models 
with smallest cell sizes $R_x,R_y\simlt$0.5 can be ruled out at 95\% 
confidence.
As the second cell size $R_y$ is increased, the curves shift to the left of 
the $T^2$-models and we can rule out $T^1$-models for 
$R_x\simlt$0.5 at a similar confidence level, see Figure~2B 
(lower plot). In this plot, we show the cumulative probability distribution of 
$S_{\circ}$ obtained from Monte Carlo simulations for the cell sizes 
($R_x,R_y,R_z$) = (0.5,3,3), (0.6,3,3), (0.7,3,3) and (3,3,3). 

A more complete picture of the cell size limits is obtained when we construct a
two-dimensional grid for different values of the cell sizes ($R_x,R_y,R_z$) 
with $R_z=3.0$ and $0.2<R_x,R_y<3.0$ (see Figure~3). 
The thin-shaded, thick-shaded and grey regions correspond,
respectively, to the models ruled out at 68\%, 90\% and 95\%  
confidence.
Notice in this plot that all contours are almost $L$-shaped, which means that, 
to a good approximation, the level in which a model ($R_x,R_y$) is ruled out 
depends only on the {\it smallest} cell size, $R_{min} \equiv \{ R_x,R_y \}$. 
We see that $R_{min}\simgt$0.5 at 95\% confidence.

\section{CONCLUSIONS}

We have shown that the $COBE$/DMR maps have the ability to discriminate and 
rule out $T^1$ and $T^2$ topological models. 
We have presented a new statistic to study these anisotropic models which 
quantifies the ``smallness" of a sky map in a single number, $S_{\circ}$, 
which is independent of the cell orientation, is easy to work with and 
is easy to interpret.

From the $COBE$/DMR data, we obtain a lower limit for $T^1$- and $T^2$-models 
of $R_x\simgt$0.5, which corresponds to a cell size with smallest dimension
of $L$=3000$h^{-1}$Mpc.
This limit is at 95\% confidence and assumes $n$=1. 
Since the topology is interesting only if the cell size is considerably
smaller than the horizon, so that it can (at least in principle) be directly 
observed, these models lose most of their appeal. 
Since the cubic $T^3$ case has already been ruled out as an interesting 
cosmological model (see {\eg} dOCS95), and $T^1$- and $T^2$-models for small 
cell sizes are ruled out, this means that {\it all} toroidal models (cubes and
rectangles) are ruled out as interesting cosmological models.

\bigskip

We would like to thank Jon Aymon and Al Kogut for the help with the $COBE$
library subroutines and Douglas Scott, Daniel Stevens and Max Tegmark for many 
useful comments and help with the manuscript. 
AC acknowledges SCT-PR/CNPq Conselho Nacional de Desenvolvimento Cient\'\i fi-
co e Tecnol\'{o}gico for her financial support under process No.~201330/93-8(NV). 
AS was supported in part by the Russian Foundation for Basic Research,
under grant No.~93-02-3631, and by the Russian Research Project 
"Cosmomicrophysics".
AC and AS were also supported in part by the National Science Foundation,
under grant No.~PHY94-07194, during their visit to ITP, USCB and
participation in the Conference on CMB Fluctuations (Santa Barbara,
22-24 February 1995) where this project was begun.
This work was also supported in part by the Director, Office of Energy 
Research, Office of High Energy and Nuclear Physics, Division of High Energy 
Physics of the U.S. Department of Energy under contract No.~DE-AC03-76SF00098.

\clearpage

\centerline{\bf REFERENCES}
\bigskip
\noindent
Bond, J.R. \& Efsthatiou, G.P. 1987, MNRAS, 226, 655.\\
Broadhurst, T.J., Ellis, R.S., Koo, D.C. \& Szalay, A.S. 1990, Nature, 343,726.\\
de Oliveira-Costa, A. \& Smoot, G.F. 1995, ApJ, 448, 447 (``dOCS95"). \\
Ellis, G.F.R. \& Schreiber, G. 1986, Phys. Lett. A, 115, 97.\\
Fang, L.Z. 1993, Mod. Phys. Lett. A, 8, 2615. \\ 
Fang, L.Z. \& Houjun, M. 1987, Mod. Phys. Lett. A, 2, 229. \\ 
Fang, L.Z. \& Sato, H. 1985, Gen. Rel. and Grav., 17, 1117. \\
Jing, Y.P. \& Fang, L.Z. 1994, Phys. Rev. Lett., 73, 1882. \\
Lineweaver, C. et al. 1994, ApJ, 436, 452. \\
Peebles, P.J.E. 1982, ApJ, 263, L1. \\
Scott, D., Srednicki, M. \& White, M. 1994, ApJ, 421, L5. \\
Smoot, G.F. et al. 1992, ApJ, 396, L1. \\
Sokolov, I.Y. 1993, JETP Lett., 57, 617.\\
Starobinsky, A.A. 1993, JETP Lett., 57, 622 (``S93").\\
Stevens, D., Scott, D. \& Silk, J. 1993, Phys. Rev. Lett., 71, 20. \\
Zel'dovich, Ya B. 1973, Comm. Astrophys. Space Sci., 5, 169.\\
Zel'dovich, Ya B. \& Starobinsky, A.A. 1984, Sov. Astron. Lett., 10, 135.\\

\clearpage

\centerline{\bf FIGURE CAPTIONS}

\bigskip

\noindent
Figure~1:
Simulated sky maps for the $T^1$- and $T^2$-models and their $S$-maps.
(A) $T^1$-model with dimensions ($R_x,R_y,R_z$) = (3,3,0.3);
(B) $T^2$-model with dimensions ($R_x,R_y,R_z$) = (0.3,0.3,3); 
(C) $S$-map of the $T^1$-model shown in ($A$); 
(D) $S$-map of the $T^2$-model shown in ($B$).

\bigskip

\noindent 
Figure~2: 
Cumulative probability distribution of $S_{\circ}$ for $T^1$- and 
$T^2$-models obtained from Monte Carlo simulations. 
(A, upper plot) Simulations for $T^2$-universes with dimensions ($R_x,R_y,R_z$) = 
(0.5,0.5,3) or dot-dashed line, (0.6,0.6,3) or dashed line, and 
(0.7,0.7,3) or dotted line. 
(B, lower plot) Simulations for $T^1$-universes with dimensions ($R_x,R_y,R_z$) = 
(0.5,3,3) or dot-dashed line, (0.6,3,3) or dashed line, and 
(0.7,3,3) or dotted line. 
In both pictures the model ($R_x,R_y, R_z$) = (3,3,3) is represented by a solid 
line, $S_{\circ}^{DMR}$ = 2.59 (vertical straight line) and the horizontal 
solid lines indicate the confidence levels of 95\%, 90\% and 68\% 
(from top to bottom).

\bigskip

\noindent
Figure~3:
Grid of cumulative probability distributions of $S_{\circ}$ for 
$T^1$- and $T^2$-models obtained from Monte Carlo simulations.  
The thin-shaded, thick-shaded and grey regions correspond,
respectively, to the models ruled out at 68\%, 90\% and 95\% 
confidence.







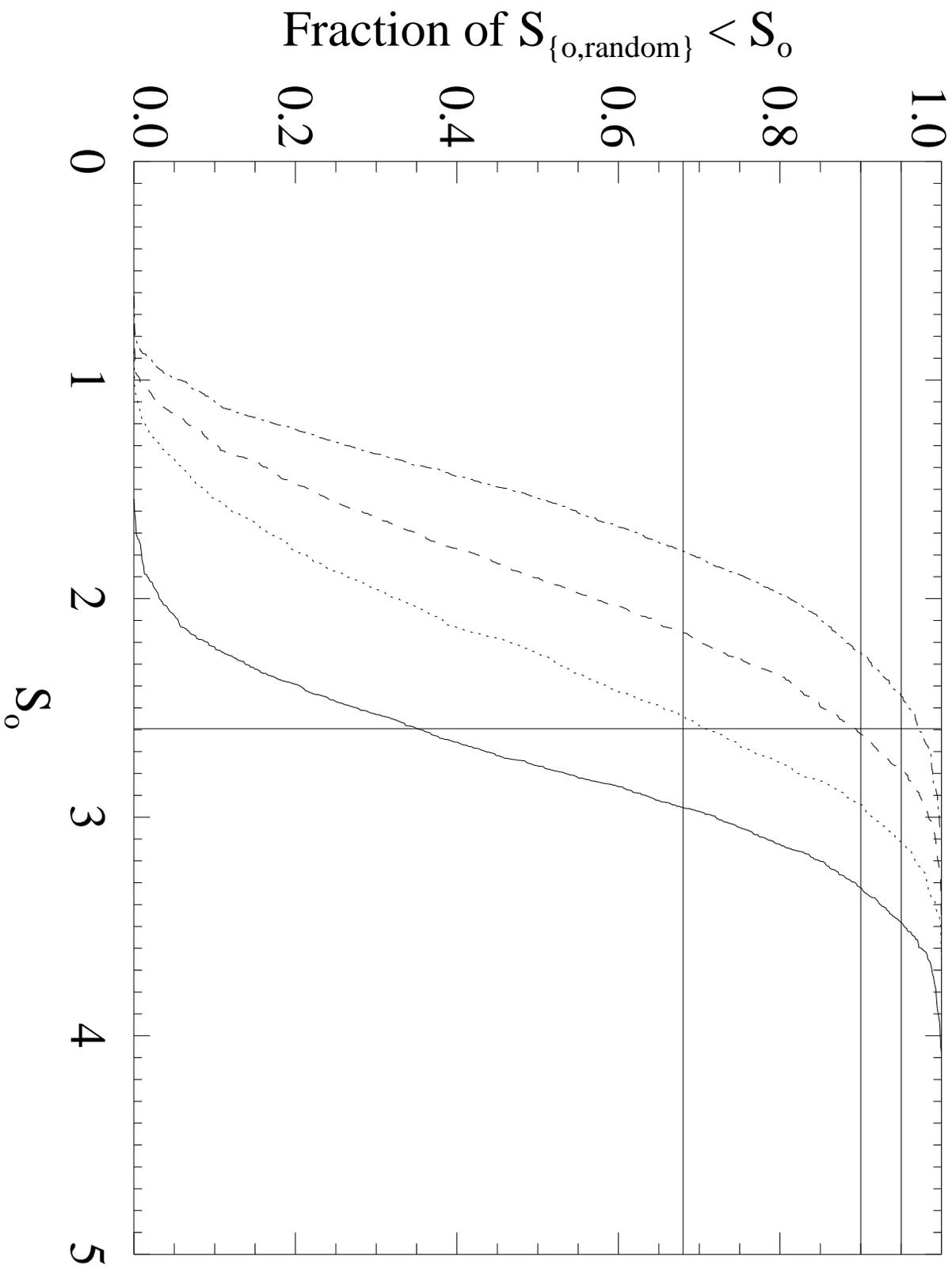
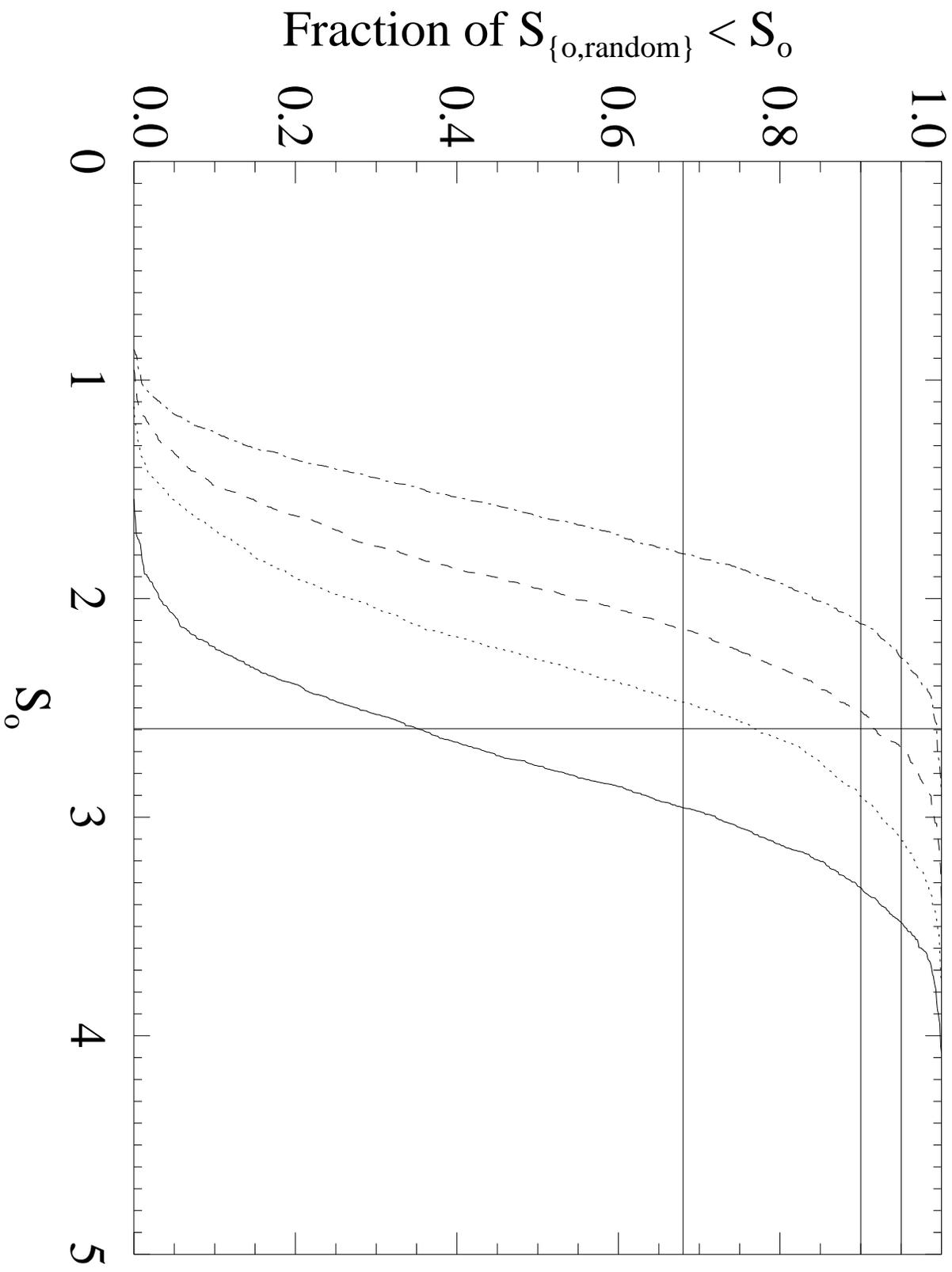
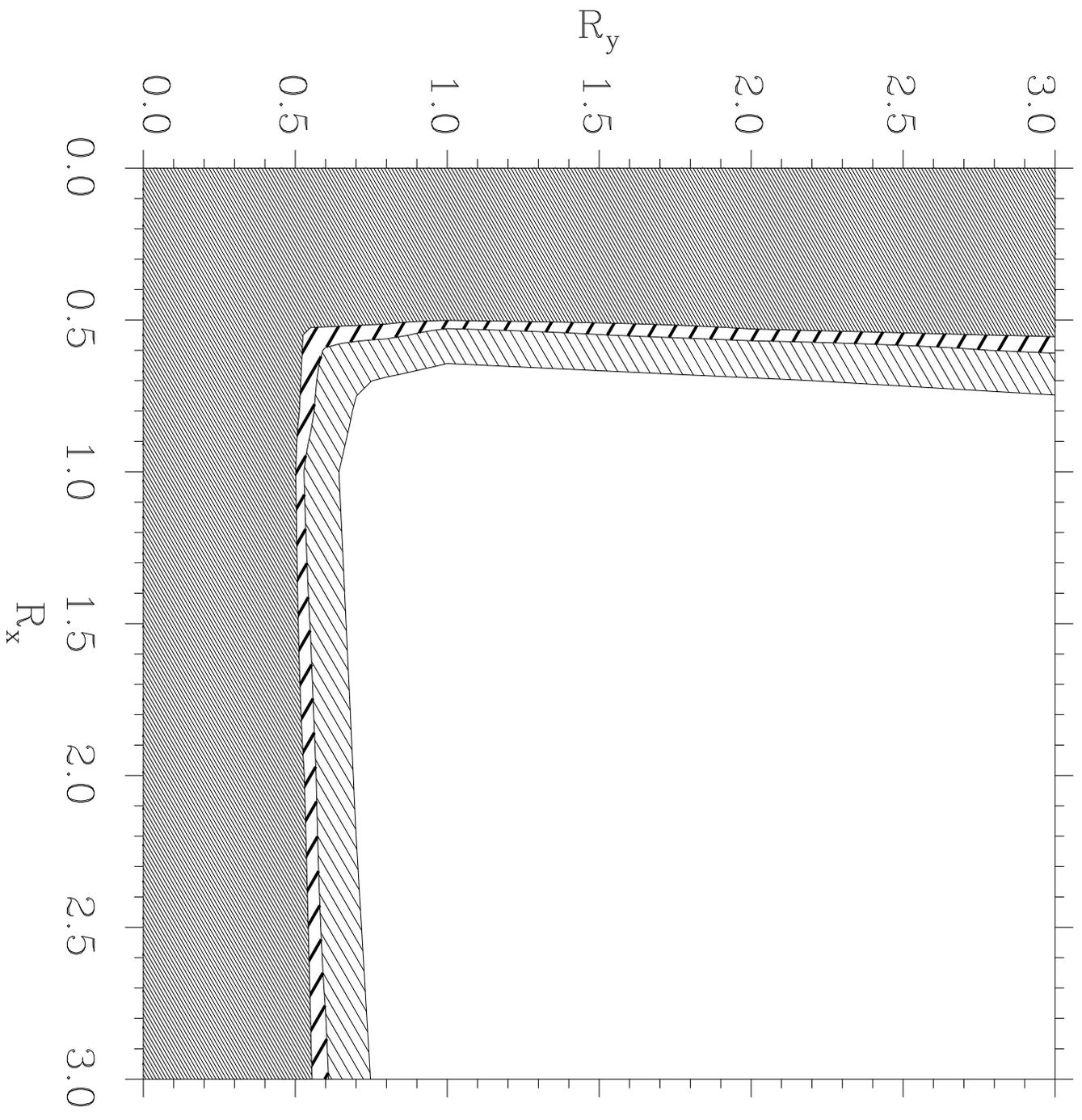
\end{document}